# Random Sequential Adsorption of Mixtures of Dimers and Monomers on a Pre-Treated Bethe Lattice

António M. R. Cadilhe*  and  Vladimir Privman**

*GCEP-Centro de Física da Universidade do Minho,
Campus de Gualtar, 4710-057 Braga, Portugal

**Center for Advanced Materials Processing, Clarkson University,
Potsdam, New York 13699-5720, USA

**Abstract:** We report studies of random sequential adsorption on the pre-patterned Bethe lattice. We consider a partially covered Bethe lattice, on which monomers and dimers deposit competitively. Analytical solutions are obtained and discussed in the context of recent efforts to use pre-patterning as a tool to improve self-assembly in micro- and nano-scale surface structure engineering.

Recently, several experimental efforts [1-16] have focused on approaches of pre-treating a surface, by imprinting microscale, and, ultimately, nanoscale, patterns, in hopes of producing new functional substrates for self-assembled structures and other applications. Kinetics of deposition of transported particles on such patterned surfaces warrants new theoretical studies. In this work, we initiate the consideration of such processes by studying random sequential adsorption, RSA, reviewed in [17,18], on the Bethe lattice [19,20].

The Bethe lattice of coordination number $z = 2, 3, \ldots$, has each site connected by bonds to $z$ nearest neighbor sites, and there are no closed loops formed by these bonds. This no-loops (no returns) property usually makes the Bethe-lattice results correspond to high-dimensional behavior. However, this property also allows exact solvability for some models, for general connectivity values $z$. We will consider an infinite lattice and ignore any boundary-site effects. The cases $z = 2, 3$ are illustrated in Figure 1.

An important property of a loopless lattice, shared with the one-dimensional lattice for which $z = 2$, is that the sites of each $k$-site cluster are connected by exactly $k-1$ internal bonds. This property is well known and established by induction: each new site can only be connected to a single existing cluster site, by a single bond, because loops are not possible. Furthermore, the number of bonds shared by the $z$ cluster sites and the nearest neighbor sites immediately outside this cluster, is $kz - 2(k-1)$. Here the first term is the total number of neighbors seen by all the $k$ cluster sites, while the term $2(k-1)$ subtracts the number of neighbors internal to the cluster. Notice that for loopless lattices, this number of shared bonds is not affected by the configuration of the cluster.

We consider RSA of dimers, arriving at the rate $D$ per unit time, $t$, and per each nearest-neighbor pair of sites, and depositing provided both of these sites are empty. We also allow deposition of monomers, arriving at the rate $M$ per unit time and per each lattice site, and depositing only at those sites that are empty. Arriving particles that cannot be deposited are discarded. A standard approach to solving RSA problems [21] involves the consideration of the probabilities that a randomly chosen $k$-site connected cluster is empty, $P_k(t)$. The occupancy of the neighbor sites of the cluster is not accounted for in defining this probability. Thus, some of the clusters that are empty can be parts of larger empty clusters. With these definitions, we can write the rate equations for the empty-cluster probabilities,

$$-\frac{dP_k}{dt} = MkP_k + D\big[(k-1)P_k + (zk - 2k + 2)P_{k+1}\big], \qquad (1)$$

where the term proportional to $P_{k+1}$ corresponds to those dimers that deposit on a pair of sites only one of which is in the considered $k$-site cluster.

Let us now define the relative deposition attempt rates,

$$\alpha = \frac{D}{M+D}, \tag{2}$$

$$\beta = \frac{M}{M+D} = 1-\alpha. \tag{3}$$

We will also rescale the time variable to correspond to the dimensionless combination

$$t_{\text{new}} = (M+D)t_{\text{old}}. \tag{4}$$

The rate equations then read

$$-\frac{dP_k}{dt} = \beta k P_k + \alpha\left[(k-1)P_k + (zk-2k+2)P_{k+1}\right], \tag{5}$$

and they can be solved by the ansatz

$$P_k(t) = f(t)\left[g(t)\right]^k, \tag{6}$$

provided the initial conditions, at $t=0$, are also of the form (6). Indeed, the functions entering (6) can be determined from

$$-\frac{dg}{dt} = g + \alpha(z-2)g^2, \tag{7}$$

$$\frac{d\ln f}{dt} = \alpha(1-2g). \tag{8}$$

We will now address the question of whether the initial conditions for pre-patterned substrates are of the form (6). One way to pre-treat the substrate is by randomly blocking some sites for deposition, with the initial fraction of the remaining empty sites given by $0 \le \rho \le 1$. In a sense, this is equivalent to rapid initial deposition of monomers, followed by mixture deposition. Deposition processes of mixtures, as well as deposition on finite-size lattices and on randomly covered lattices, have been considered in the literature, e.g., [17-19,22-30]. The ansatz (6) applies in this case, with

$$f(0) = 1, \tag{9}$$

$$g(0) = 1-\rho. \tag{10}$$

We find

$$P_k(t) = (1-\rho)^k e^{(\alpha-k)t}\left[1+\alpha(z-2)(1-\rho)(1-e^{-t})\right]^{-\left(k+\frac{2}{z-2}\right)}. \qquad (11)$$

The coverage, $\theta$, defined as the fraction of sites covered by the deposited dimers and monomers, is given by

$$\theta(t) = 1 - P_1(t). \qquad (12)$$

We get

$$\theta(t) = 1 - (1-\rho)e^{-\beta t}\left[1+\alpha(z-2)(1-\rho)(1-e^{-t})\right]^{-\frac{z}{z-2}}. \qquad (13)$$

This expression has interesting limiting properties as $z \to 2$, the one-dimensional case, and for $\alpha = 1$, corresponding to the dimer-only deposition and formation of a typical RSA jammed state with the final coverage less than 1, whereas $\theta(\infty) = 1$ for any $0 \leq \alpha < 1$. However, we will not consider these details here.

Instead, let us focus on the issue of pre-treating the substrate. The random coverage, just considered, is an important case that finds experimental realizations [31-34]. However, to actually control the particle deposition for self-assembly, we have to consider much more restrictive situations, when the substrate is pre-treated in such a way that particle deposition is possible only in specific locations tailored to their sizes. Let us, therefore, consider the initial conditions corresponding to only voids of size 1 and 2 left uncovered.

We will denote by $S$ the initial fraction of sites that are single-site voids, and by $T$ the initial fraction of sites that, pair-wise, form two-site voids. Then, we have

$$P_1(0) = S + T, \qquad (14)$$

$$P_2(0) = \frac{T}{z}, \qquad (15)$$

and

$$P_{k>2}(t \geq 0) \equiv 0, \qquad (16)$$

while the $k = 1, 2$ rate equations (5) yield

$$-\frac{dP_1}{dt} = \beta P_1 + \alpha z P_2, \qquad (17)$$

$$-\frac{dP_2}{dt} = (2-\alpha)P_2. \tag{18}$$

For definiteness, let us consider the solution for the total coverage fraction, still given by $\theta = 1 - P_1$,

$$\theta(t) = 1 - (S + \beta T + \alpha T e^{-t})e^{-\beta t}. \tag{19}$$

With the initially covered fraction of sites $\rho = 1 - S - T$, this can be rewritten as

$$\theta(t) = 1 - (1-\rho)e^{-\beta t}\left[1 - \frac{\alpha T}{1-\rho}(1 - e^{-t})\right]. \tag{20}$$

There is certain similarity between the functional forms of the time-dependence in (13) and (20). However, not surprisingly (20) is not sensitive to the coordination number of the lattice.

In summary, we obtained exact solutions for deposition processes on the Bethe lattice with the substrate pre-treated by either random blocking of a fraction of sites or by a more specific preparation that allows only for monomer and dimer landing slots. The time-dependent surface coverage was calculated in terms of the problem parameters. We hope that the functional forms found can be of use in fitting experimental data as well as for guiding numerical studies for more realistic geometries for which analytical solutions are not possible.

This research has been supported by Fundação para a Ciência e a Tecnologia (grant POCTI/CTM/41574/2001 – Computational Nanophysics) and the US National Science Foundation (grant DMR-0102644). One of the authors (A.M.R.C.) acknowledges the hospitality of the Colloid Science Group at Clarkson University.

**Figure**

**Figure 1:** Examples of Bethe lattices for (a) $z = 2$ and (b) $z = 3$, with the origin at the central site C.

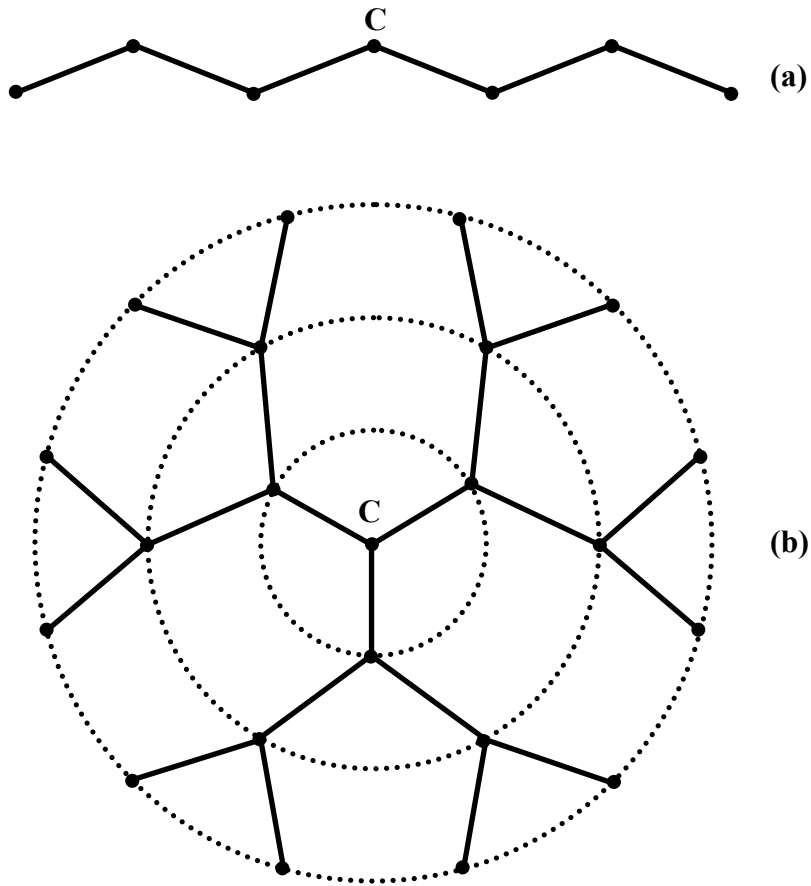

# Table of Contents Graphic

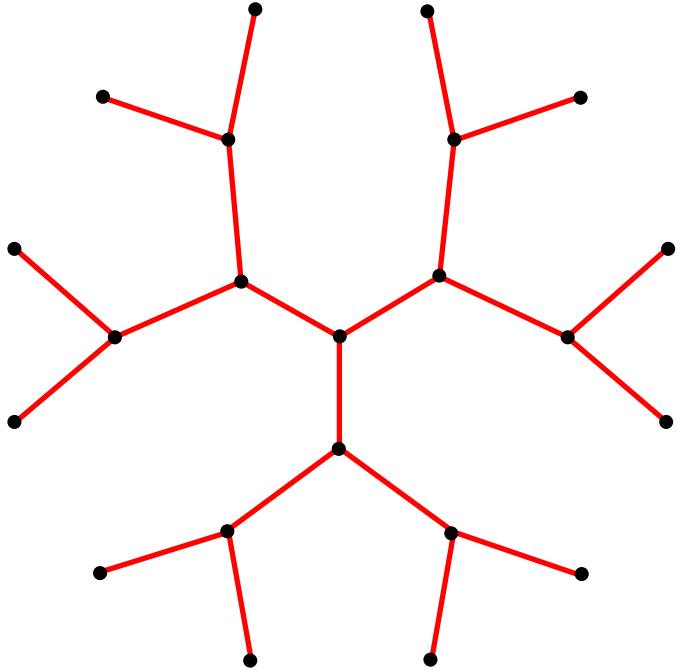